\documentclass[floatfix,twocolumn,showpacs,preprintnumbers,amsmath,amssymb,prr,superscriptaddress,longbibliography, nofootinbib]{revtex4-1}

\usepackage{color}    
\usepackage{graphicx}
\usepackage{dcolumn}
\usepackage{bm}
\usepackage{subfigure}
\usepackage{amssymb}
\usepackage{multirow}
\usepackage{natbib}
    \usepackage{amsmath}
\graphicspath{{plots/}}
\usepackage{hyperref}
\hypersetup{
    colorlinks = true,
    citecolor = blue
}
\usepackage[utf8]{inputenc}
\usepackage[T1]{fontenc}
\usepackage[normalem]{ulem}
\usepackage{dirtytalk}
\usepackage{appendix}
\usepackage{tabularx}
\newcommand{\eps}[0]{\varepsilon}	

\renewcommand{\vec}[1]{\mathbf{#1}}

\begin{document}

\title{Shear Viscosity in Two-Dimensional Dipole Systems}
\author{N.~E.~Djienbekov}
\author{N.~Kh.~Bastykova}
\author{A.~M.~Bekbussyn}
\author{T. S.~Ramazanov}
\author{S. K.~Kodanova}
\email[]{kodanova@physics.kz}

\affiliation{
Institute for Experimental and Theoretical Physics, Al-Farabi Kazakh National University, 71 Al-Farabi ave.,
 050040 Almaty, Kazakhstan}
 
\begin{abstract}

    The results of modeling shear flows in classical two-dimensional dipole systems are presented. We used the method of non-equilibrium molecular dynamics to calculate the viscosity at various shear rates. The coefficients of shear viscosity are given in the limit of low shear rates for various regimes of interparticle correlation from a weakly correlated  gaseous state to a strongly non-ideal liquid state near the crystallization point. The calculations were carried out for bare (unscreened) dipole systems, as well as for dipole systems in a polarizable medium that provide screening of the dipole–dipole interaction. The effect of shear thinning in 2D dipole systems is reported at small values of the coupling parameter. In addition, it is shown that dipole systems can become both less and more viscous due to the presence of a screening medium, depending on the degree of interparticle correlation. The optimal simulation parameters are discussed within the framework of the method of nonequilibrium molecular dynamics for determining the shear viscosity of two-dimensional dipole systems. Moreover, we present a simple  fitting curve which provides  universal scaling law for both bare dipole - dipole interaction and screened dipole-dipole interaction.

\end{abstract}
\maketitle
\section{INTRODUCTION}
Two-dimensional systems governed by a repulsive dipole-dipole pair interaction are relevant for various systems.
For example, the repulsive dipole-dipole interaction is used to describe two-dimensional colloidal systems \cite{Zanghellini, PhysRevLett.93.255703, Pretti}. 
In complex plasmas, the interaction between charged dust particles can be modified due to external fields and fluxes of ions and electrons \cite{doi:10.1063/1.3112703, PhysRevLett.116.125001, Khrapak, PhysRevResearch.3.043187, cpp_15_zm, doi:10.1063/1.4922908, Sundar_2020, Sundar_2020_njp}.
It was shown that a repulsive dipole-dipole interaction is realized in complex plamsas at certain conditions \cite{Cooper, Kompaneets_pre, Sukhinin, Lapenta, Khrapak, Sukhinin_2018, Aldakul, Gabdulin, Kodanova}.
Furthermore, a system of polar molecules \cite{Lemeshko} and a dipolelike excitonic phase state (created by bound electron-hole excitons) can be described using  a model of classical 2D system of dipoles \cite{Golden1, Golden2}.

Aforementioned examples have motivated studies of various properties of classical two-dimensional  systems using the repulsive dipole-dipole potential \cite{Khrapak3}. 
For example, Khrapak et al \cite{Khrapak} investigated thermodynamic and dynamic properties of a classical 2D system of dipoles.
Earlier, the characteristic  oscillation modes of particles in the 2D dipole system were analyzed  by Golden et al \cite{Golden1, Golden2}. 
In Refs. \cite{Golden1, Golden2}, it was  demonstrated  that a dipolelike excitonic phase state created by bound electron-hole excitons in semiconductors can be
described using model of a classical 2D system of repulsive dipoles. 
These works on oscillation modes in 2D dipole systems were continued  by the study of
the dumping of the transverse excitations in the long wave length domain \cite{Mistryukova, PhysRevLett.125.125501}.  
More recently, Aldakul et al \cite{Aldakul} investigated melting, freezing, and the liquid-crystal phase transition point  of classical 2D dipole systems. 
In this work, we extend these studies of 2D dipole systems by modeling shear viscosity and  shear flows in classical 2D systems with repulsive dipole interaction across coupling regimes. 

In addition to a standard dipole-dipole interaction, in this work we use screened dipole-dipole interaction. In the latter case screening can be due to a polarizable medium surrounding 2D dipole system \cite{Sukhinin_2018, Aldakul, Ramazanov}. For example, regarding aforementioned a dipolelike excitonic phase state, it was recently shown that screening due to excess charges modifies electron-hole excitons \cite{Tiene}. In complex plasmas, the stream of ions creates a focused ion cloud near a charged dust particle in downstream direction due to attraction of ions by a negative charge of a dust particle and  the inelastic collision of ions with atoms \cite{PhysRevE.95.063207, PhysRevE.79.036404}. The focused ion cloud together with the charged dust particle create an compound particle with zero total charge and non-zero dipole moment \cite{Sukhinin_2018}. Additionally, hot electrons---with the electron Debye length much larger than both the ion Debye length and the size of the compound particle--- provide screening of ion and dust particle charges at long distance \cite{Sukhinin_2018, Moldabekov_ARXIV}. This leads to the formation of the screened dipole-dipole interaction between compound particles. The impact of screening on the structural properties, oscillation modes, and thermodynamic characteristics of 2D dipole systems has been discussed in Ref. \cite{Aldakul}. 

To compute the shear viscosity of 2D systems one can use the reverse nonequilibrium molecular dynamics method (NEMD)\cite{Muller}, \cite{Sanbonmatsu}-\cite{Hartmann}.
This method was used previously to investigate shear flows in classical 2D Yukawa systems \cite{Hartmann}.
It was shown that  the NEMD  allows to determine shear viscosity in a good agreement with experimental observation \cite{Donko}.
Moreover, the NEMD allows one to study a non-Newtonian fluid behavior,  i.e., when shear viscosity vary with the velocity gradient. One of the peculiar properties of non-Newtonian fluids is  decrease of the viscosity  as shear is increased. This effect is referred to as  shear thinning. For example, following original studies on simple liquids by Evans et al \cite{Evans}, such behavior has been reported in dusty plasmas \cite{Liu}. 
Additionally, we compere results from the NEMD simulations with the data for the shear viscosity computed using the Green-Kubo relation connecting the shear viscosity and the shear stress autocorrelation function.

The paper is organized as the following: In Sec. \ref{s:pot} we present the used pair interaction potentials. In Sec. \ref{s:2} we discuss the computation method and provide simulations details.
The results are presented in Sec. \ref{s:3}. The paper is concluded by summarizing main findings. 

\section{Bare and screened dipole-dipole interactions}\label{s:pot}

In this work, we present the results of the NEMD simulations of 2D systems with the bare dipole-dipole interaction potential: 
\begin{equation}\label{eq:bare}
\beta V(r)=\frac{\Gamma_D}{r^3},
\end{equation}
and with the screened dipole-dipole interaction \cite{Aldakul, Ramazanov}:
\begin{equation}\label{eq:screened}
\beta V(r)=\frac{\Gamma_D}{r^3}(1+\kappa r)\exp(-\kappa r),
\end{equation}
where $r$ is in the units of the mean inter-particle distance, $\beta=1/(k_BT)$ is the inverse value of a thermal energy,  $\kappa$ is screening length, and $\Gamma_D$ is the parameter characterizing coupling (correlation) strength \cite{Golden1, Golden2}.

The bare repulsive dipole-dipole pair interaction potential (\ref{eq:bare}) has been used to model two-dimensional colloidal systems \cite{Zanghellini, PhysRevLett.93.255703, Pretti} and  dipolelike excitonic phase state of bound electron-hole excitons in semiconductors \cite{Golden1, Golden2}.
The screened repulsive dipole-dipole pair interaction Eq.~(\ref{eq:screened}) provides description of dipole-dipole interaction in the presence of highly mobile polarizable background such as electrons in complex plasmas \cite{Sukhinin_2018, Lapenta, Ramazanov} and electrolyte screening field of charged colloids \cite{Everts2021-ya}.
% For the equations of motion, we need to find a force that is equal to\begin{equation}
% \bold{F_{ij}}=-\bold{\nabla}U=\frac{\exp(-k_sr)}{4\pi\epsilon_0r^4}(3+3k_sr+k_s^2r^2)\hat{\bold{r_i_j}},
% \end{equation}
% where $\bold{\hat{r_i_j}}=\bold{r}/r$, is a unit vector between dipoles $i$ and $j$. 

The coupling parameter corresponding to the melting (crystallization) point in the 2D system with bare potential (\ref{eq:bare}) is ${\Gamma}_m\simeq  67\pm 4$ \cite{Aldakul}.
The main effect of screening is to change the pair interaction from  quasi-long-range
potential to short range potential. As the result, the liquid-crystal phase transition point shifts, e.g., to ${\Gamma}_m\simeq  86 \pm 6$ at $\kappa=1$ and to ${\Gamma}_m\simeq  163 \pm 13$ at $\kappa=2 $ \cite{Aldakul}. Naturally, we report the shear viscosity results for $\Gamma_D < \Gamma_m$.

\section{COMPUTATIONAL METHOD and SIMULATION DETAILS} \label{s:2}

\subsection{The NEMD method for generating shear rate} \label{s:2a}

Let us start with a brief description of the essence of the NEMD method for the  computation of shear viscosity.
The key is to use the definition of shear viscosity in terms of a linear relationship between momentum flux and velocity gradient {\cite{Atkins}}:
\begin{equation}\label{eq:j_v}
j_x(p_x)=-\eta\frac{\partial v_x}{\partial y},
\end{equation}
where  momentum flux per unit length $j_x$, momentum $p_x$, and
shear rate $\partial v_x/\partial y$ are considered to be induced by two oppositely directed streams along $x$ axis. 

\begin{figure} 
\includegraphics[width=8cm]{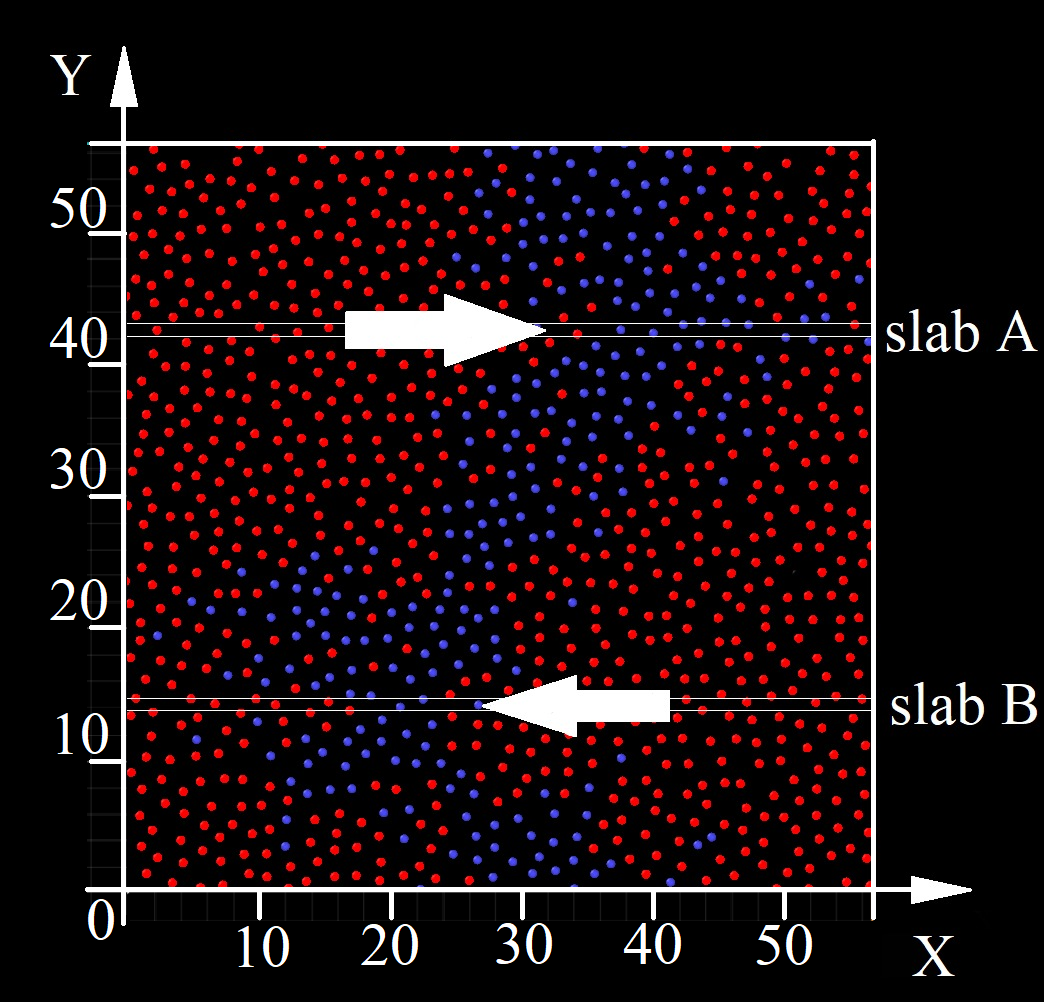}
\caption{Screenshot from a NEMD simulation after a certain amount of time after the selection of the vertical bar of particles (marked with blue), $\Gamma_D$ = 30, $\kappa$ = 2.
A horizontal shift in the position of the particles can be observed due to the presence of two oppositely directed flows generated in slabs A and B. The length is given in units of the mean-inter particle distance (see Sec. \ref{s:sim_par}).
} %\textcolor{red}{Add $y$ and $x$ axis, %indicate boundaries of slaps.}}
\label{fig:ilustration}
\end{figure}

In order to calculate shear viscosity, point-like classical particles in a simulation box with side length of $L$ are simulated with periodic boundary conditions. In the simulation box,  we define two horizontal slabs at the levels $y=L/4$ and $y=3L/4$ (see Fig. \ref{fig:ilustration}). Let us designate these slabs as A and B.
From these slabs, according to the NEMD method, the particles with the maximum and minimum values of  $v_x$ are identified  and simultaneously swapped with certain frequency (i.e. their momenta are interchanged without changing their coordinates).
In other words, the algorithm first selects the fastest particle moving to the right in the slab A  and the fastest particles moving to the left in the slab B, and, then, swaps the velocity values of these particles. As the result, the mean velocity of the particles in the slab A is directed in one direction and that of in the slab B in the opposite direction. Thus, this exchange of particle velocities conserves energy and mimics two currents flowing in opposite directions. This is  illustrated in Fig. \ref{fig:ilustration}, where a snapshot from a NEMD simulation is shown.  

To find the shear viscosity from Eq. (\ref{eq:j_v}),
first of all, the dependence of the $x$ component of the mean velocity, $v_x$, on the coordinate $y$ is computed.
Then, the value of  derivative $dv_x/dy$ in the space between two slabs is calculated using the linear regression method to find $v_x(y)$ dependence.

Second, the momentum flux is computed using the following relation: 
\begin{equation}\label{eq:flux}
j_x(p_x) = \frac{\Delta p}{2Lt},
\end{equation}
where the coefficient 2 arises due to the fact that in our case streams pass through two sides of the simulation box, $t$ is the simulation (measurement) time, and $\Delta p$ is the $x$ component of the total change in momentum as the result of swapping of velocities of particles during measurement time.
% is determined using
%  \begin{equation}
%  \Delta p=\sum_{exch}p_{x,A}-(-p_{x,B})=\sum_{exch}p_{x,A}+p_{x,B}.
%  \end{equation}

After finding the values of  shear rate $\partial v_x/\partial y$ and of  momentum flux $\Delta p$, shear viscosity is computed as 
 \begin{equation}
 \eta=\frac {\Delta p}{2tLdv_x/dy}. 
 \label{eq:9}
 \end{equation}
 
 As shown in \cite{Muller},  shear rate $\partial v_x/\partial y$ depends on the swapping frequency. 
 This means that different slope coefficients will be obtained depending how often momenta are swapped. 
 On the other hand, the momentum introduced into the system also depends on the swapping frequency. This means that, in general,  the shear viscosity of a system can depend  on the swapping frequency (i.e., shear rate).
 However, by varying the momentum exchange frequency parameter, it was found that for sufficiently rare swaps, the viscosity value is independent of the frequency of momentum exchange within statistical uncertainty \cite{Muller, Hartmann}. 
 It is this value that will be considered physically meaningful and well defined. 
 We note that the NEMD method used in this work is similar to the experimental method employing two counterpropagating laser beams to measure the shear viscosity in dusty plasmas \cite{PhysRevLett.93.155004}, where two oppositely directed flows are generated as illustrated in Fig. \ref{fig:ilustration}. 
%  More about this is discussed in the section "Calculations".

\begin{figure*}[ht]
	\begin{minipage}[c][\width]{
	   0.3\textwidth}
	   \includegraphics[width=1.2\textwidth]{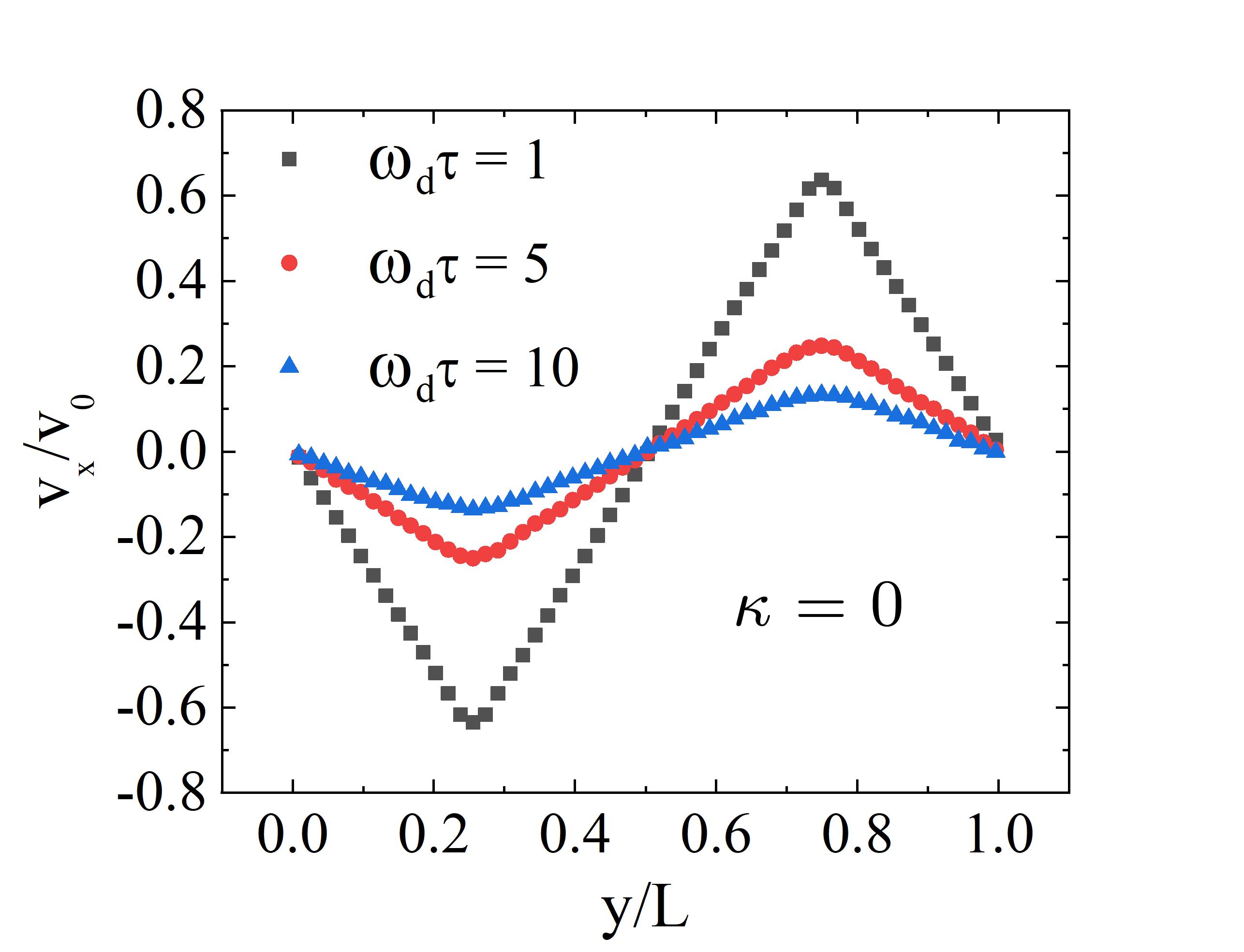}
	   %\subcaption{a) $\kappa$ = 0}
	\end{minipage}
 \hfill 	
	\begin{minipage}[c][\width]{
	   0.3\textwidth}
	   \includegraphics[width=1.2\textwidth]{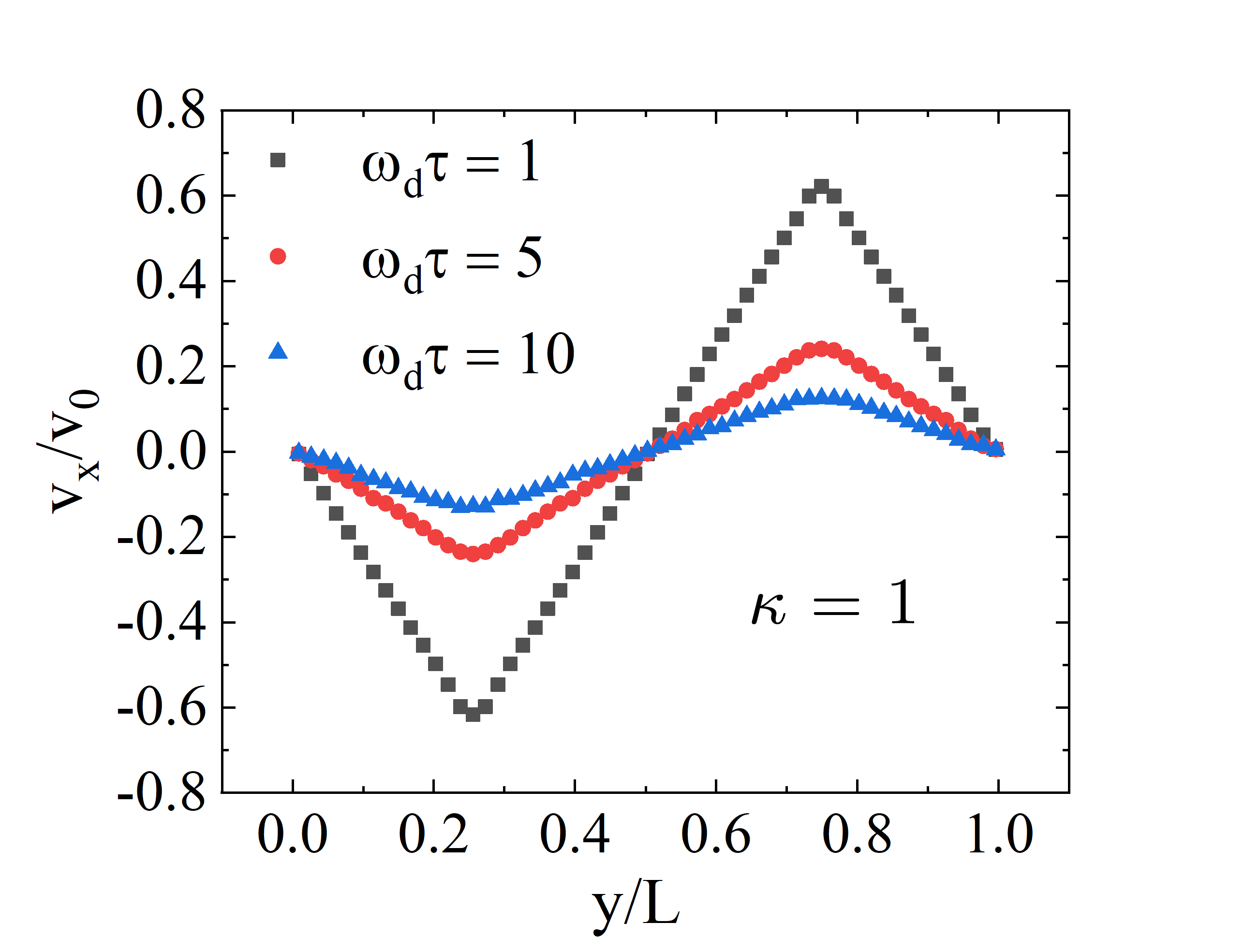}
	  % \subcaption{b) $\kappa$ = 1}
	\end{minipage}
\hfill	
	\begin{minipage}[c][\width]{
	   0.3\textwidth}
	   \includegraphics[width=1.2\textwidth]{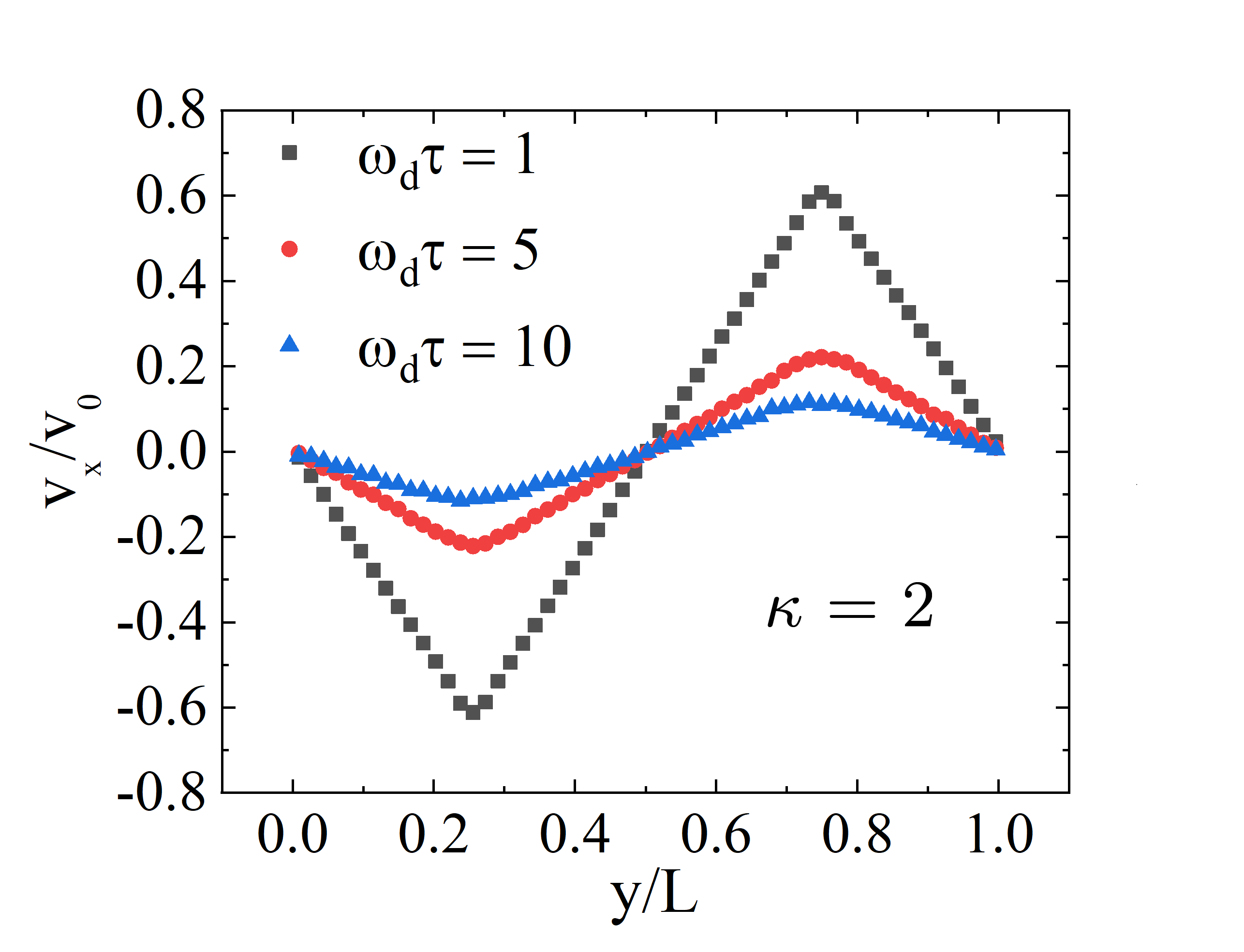}
	  % \subcaption{c) $\kappa$ = 2}
	\end{minipage}
\caption{  Velocity profiles perpendicular to the direction of flows for different screening parameters $\kappa$ at $\Gamma_D = 1$. The results are presented for different values of the period of momenta exchange between flows.
\label{fig:2}}
\end{figure*}
\begin{figure*}[ht]
	\begin{minipage}[c][\width]{
	   0.3\textwidth}
	   \includegraphics[width=1.2\textwidth]{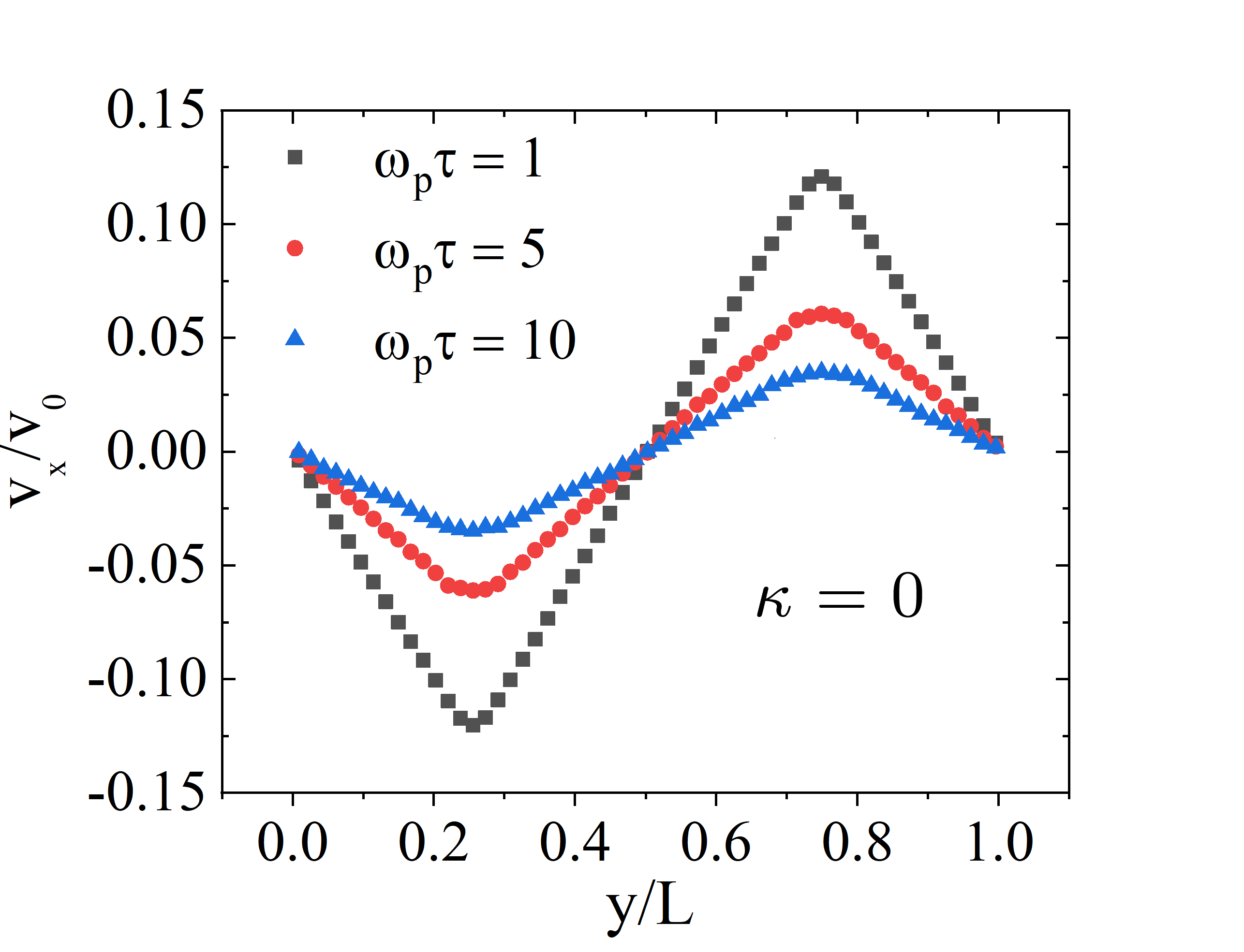}
	   %\subcaption{a) $\kappa$ = 0}
	\end{minipage}
\hfill 	
	\begin{minipage}[c][\width]{
	   0.3\textwidth}
	   \includegraphics[width=1.2\textwidth]{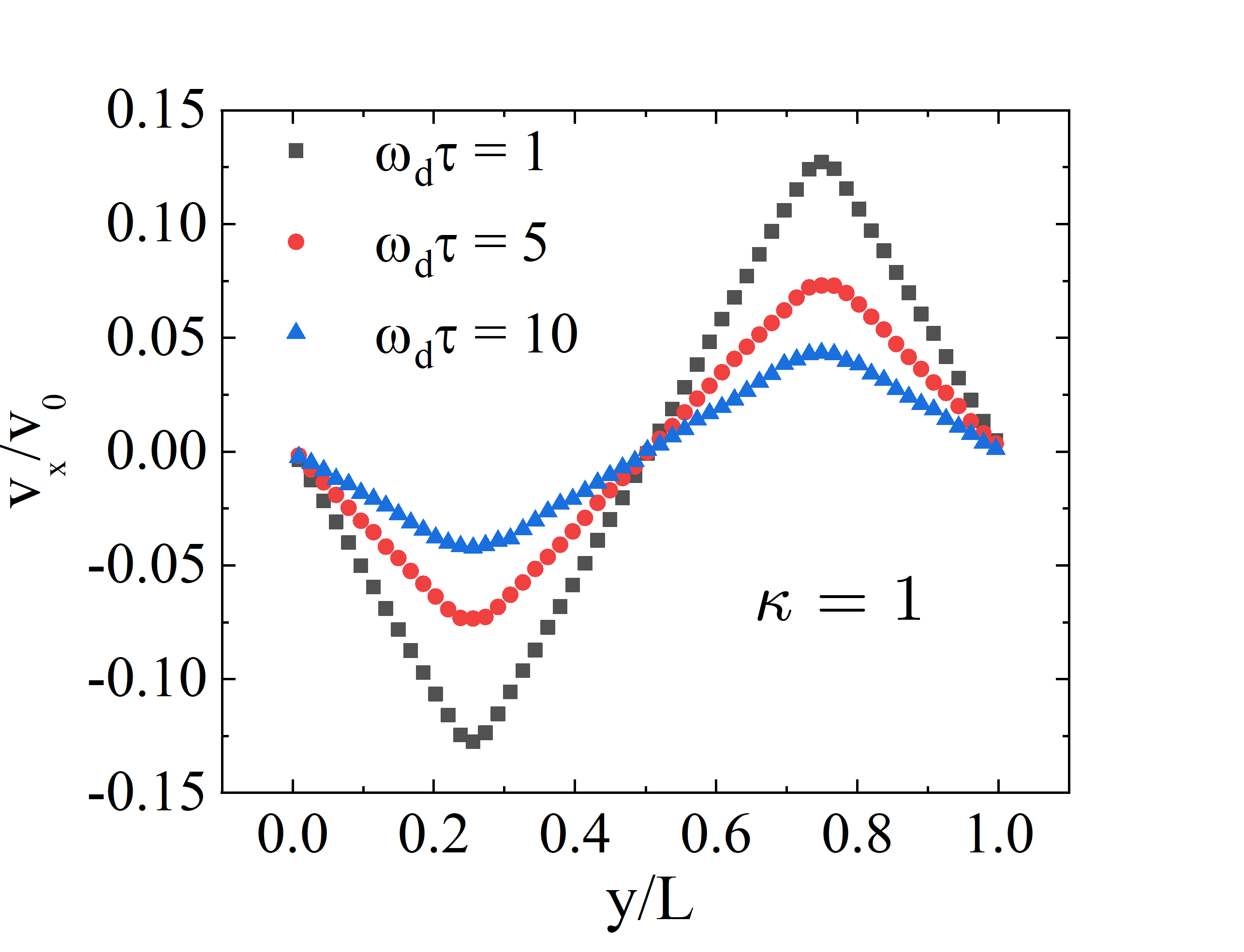}
	   %\subcaption{b) $\kappa$ = 1}
	\end{minipage}
\hfill	
	\begin{minipage}[c][\width]{
	   0.3\textwidth}
	   \includegraphics[width=1.2\textwidth]{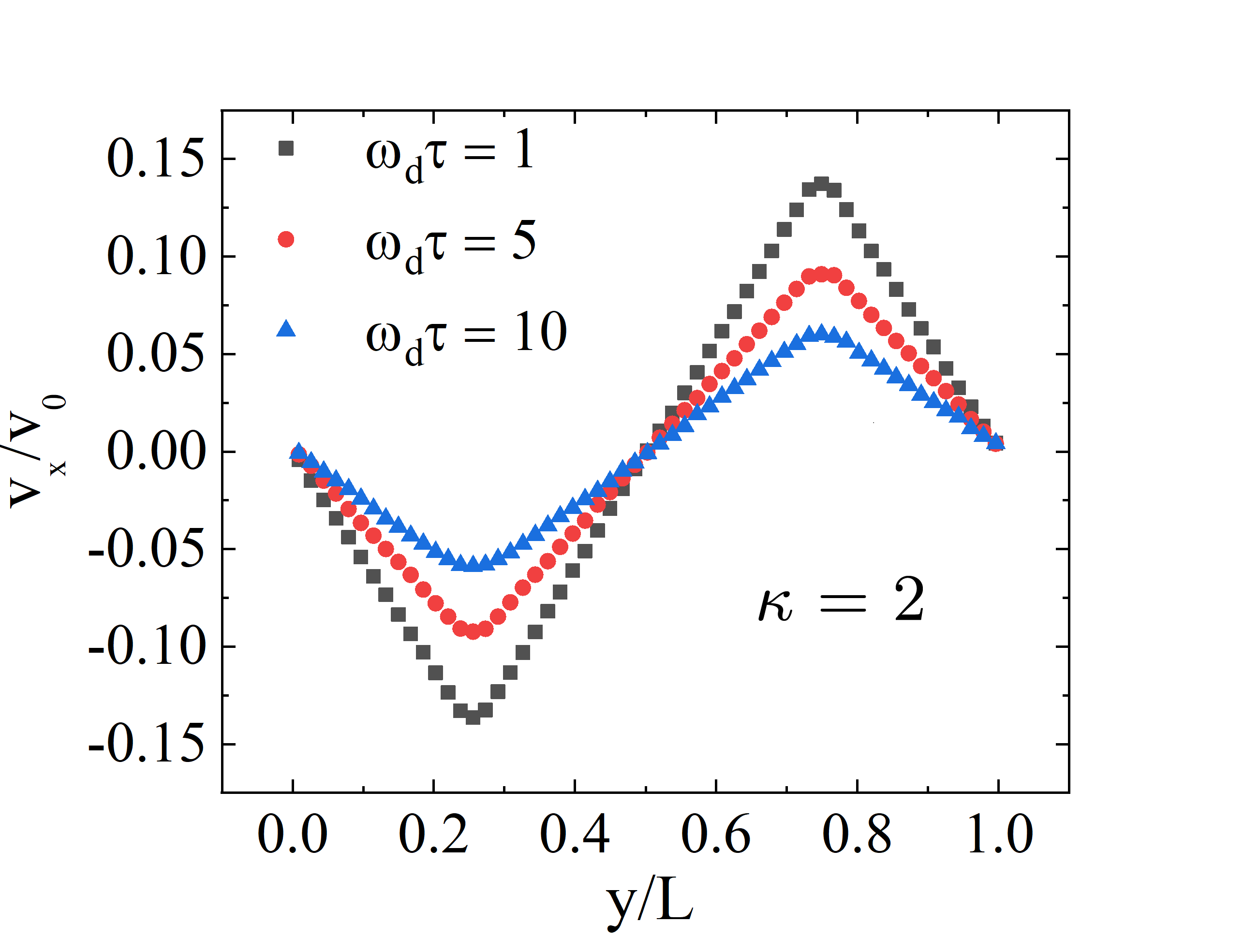}
	   %\subcaption{c) $\kappa$ = 2}
	\end{minipage}
\caption{  Velocity profiles perpendicular to the direction of flows for different screening parameters $\kappa$ at $\Gamma_D = 30$. The results are presented for different values of the period of  momenta exchange between flows. 
\label{fig:3}}
\end{figure*}

\begin{figure} 
\includegraphics[width=0.5\textwidth]{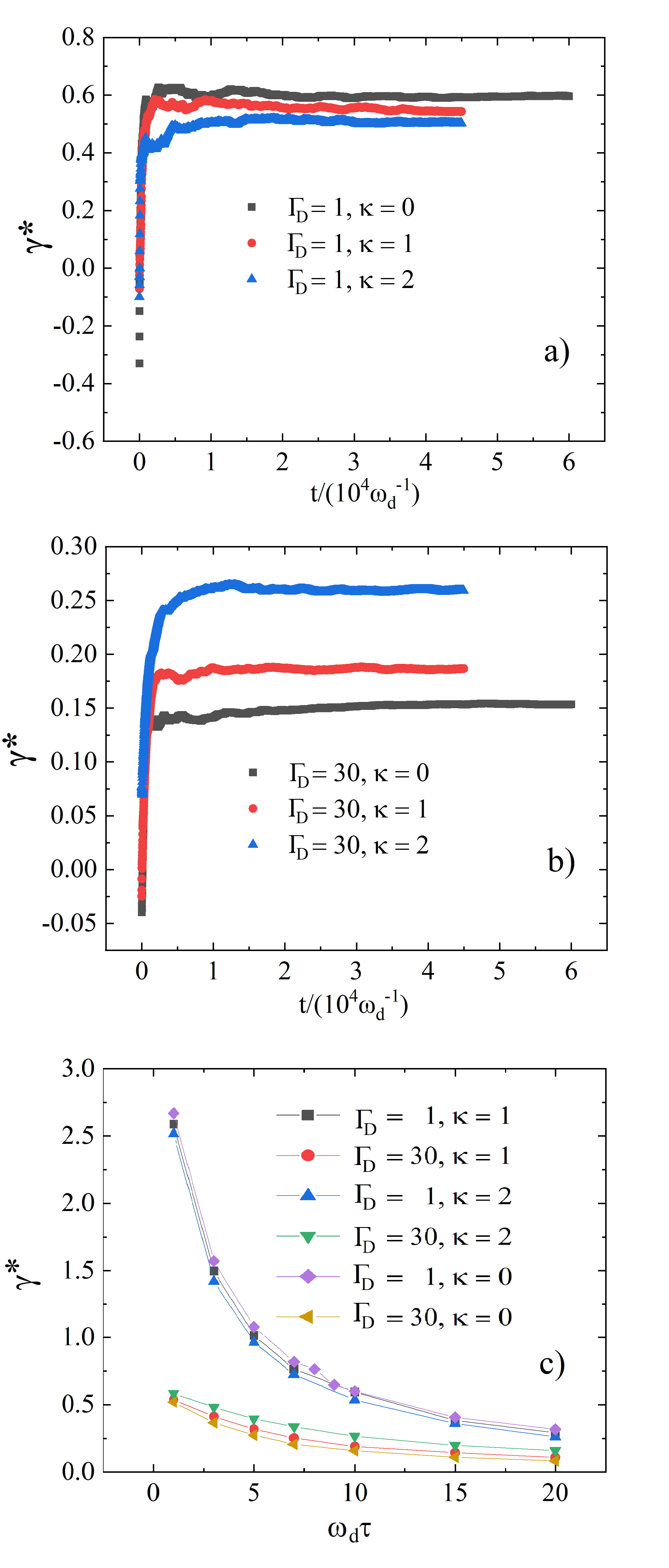}
\caption{  a)  The dependence of the shear rate on time at parameters: $\Gamma_D$ = 1; $\kappa$ = {0, 1, 2}; $\omega_d\tau$ = 10. b) The dependence of the shear rate on time at parameters: $\Gamma_D$ = 30; $\kappa$ = {0, 1, 2}; $\omega_d\tau$ = 10. c) The dependence of the shear rate on the period of momentum exchange in the slabs. 
\label{fig:4}}
\end{figure}
 
\begin{figure} 
\includegraphics[width=0.5\textwidth]{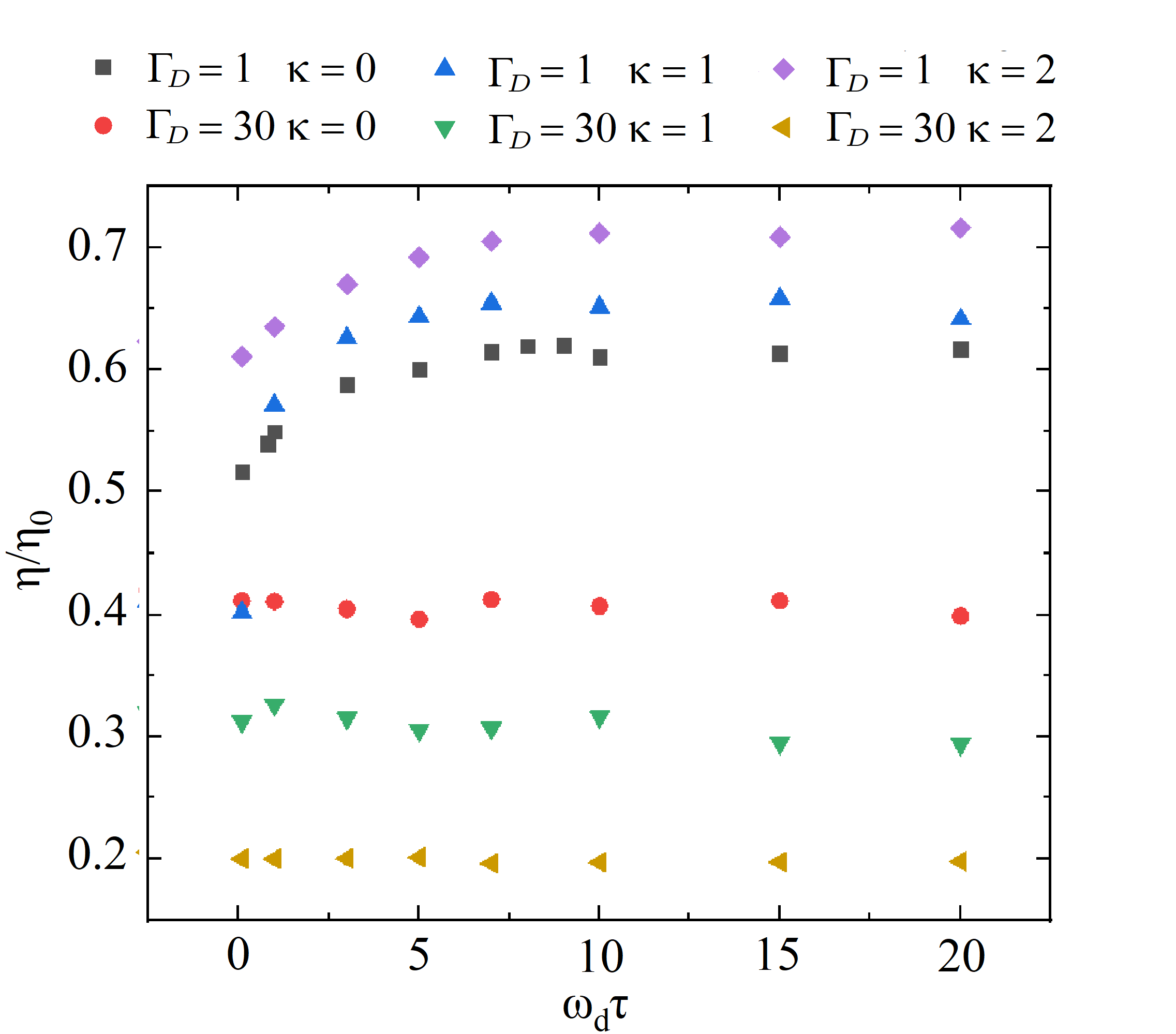}
\caption{  Shear viscosity as a function of the momentum exchange frequency for selected values of the coupling parameter $\Gamma_D$.
\label{fig:5}}
\end{figure}

\begin{figure} 
\includegraphics[width=0.5\textwidth]{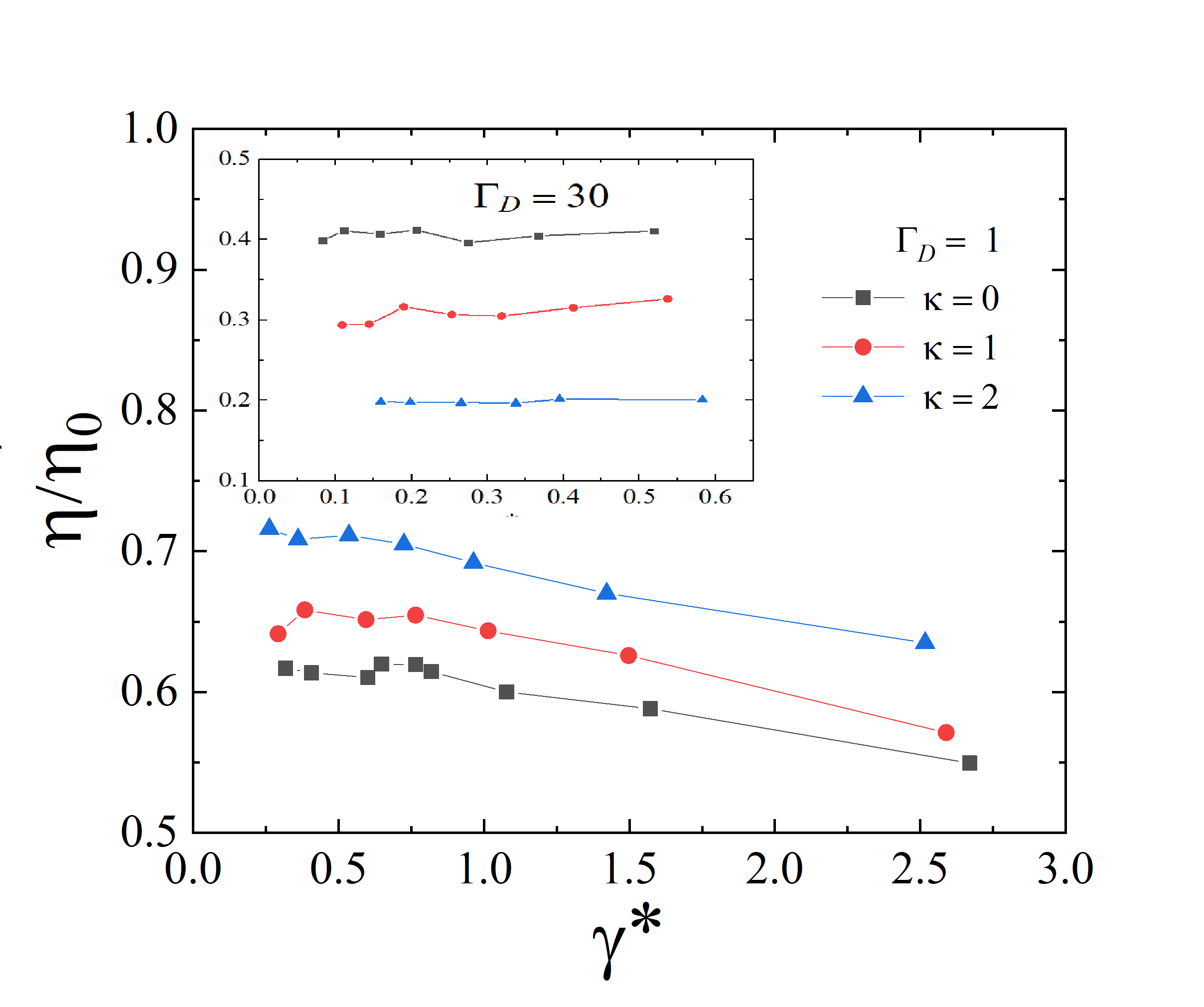}
\caption{ Shear viscosity as a function of the shear rate for selected values of the coupling parameter $\Gamma_D$.
\label{fig:6}}
\end{figure}

\begin{figure} 
\includegraphics[width=0.5\textwidth]{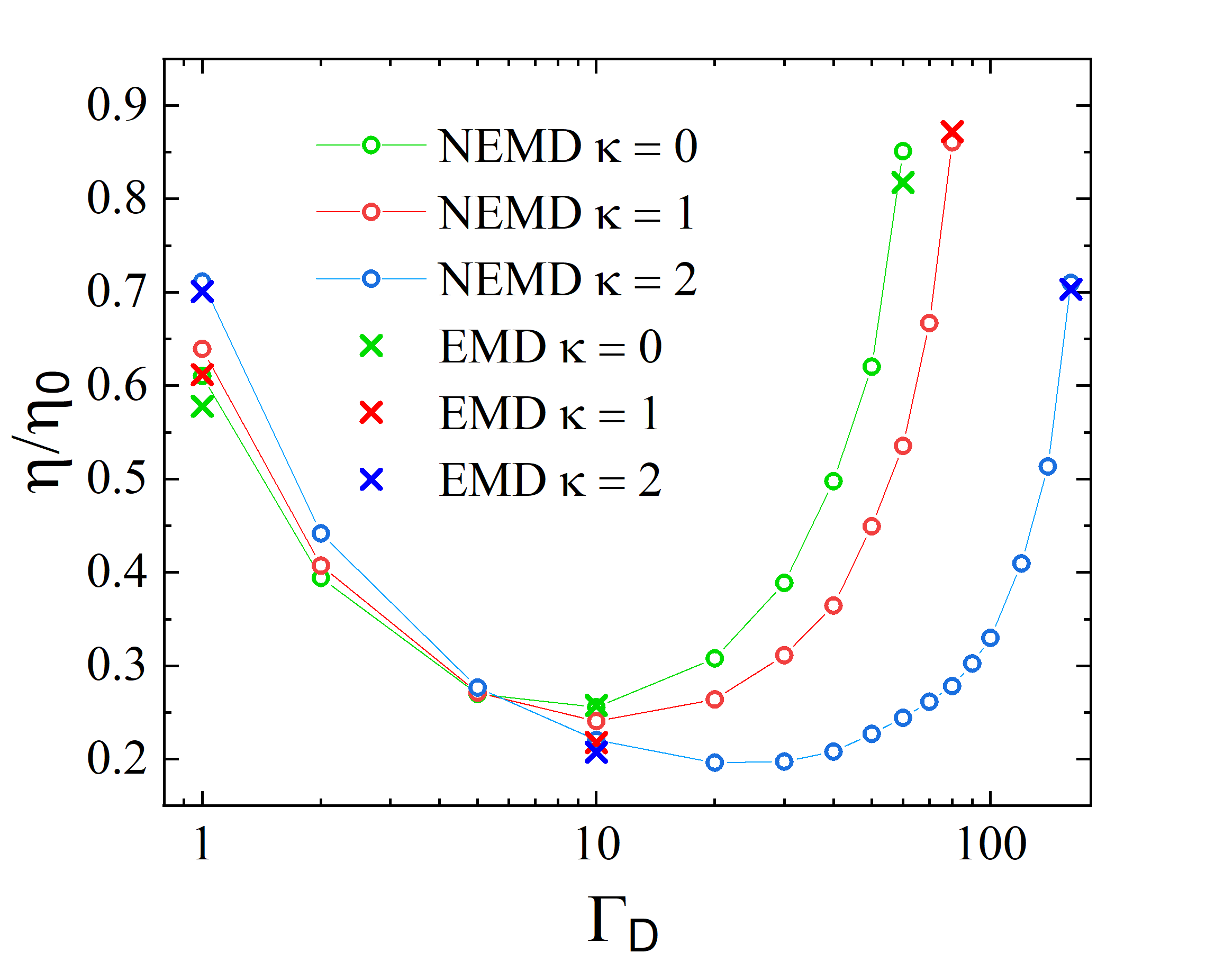}
\caption{  Shear viscosity in the limit of low shear rates from the NEMD and EMD simulations. \label{fig:7}}
\end{figure}

\subsection{Shear viscosity from equilibrium molecular dynamics} \label{s:2b}

As an extra cross check, we used equilibrium molecular dynamics to calculate the shear viscosity from the Green-Kubo relation connecting  the shear viscosity with the shear stress autocorrelation
function. We refer to this approach as equilibrium molecular dynamics method (EMD).

Within the EMD, the Green-Kubo relation for the shear viscosity  reads:
\begin{equation}\label{eq:Green-Kubo}
\eta = \frac{1}{Sk_BT}\int_{0}^{\infty} C(t) \,dt, \quad  C(t)=\left<P^{xy}(t)P^{xy}(0)\right>,
\end{equation}
where $S$ is the area of the simulation box and $P^{xy}$ is the off-diagonal element of the pressure tensor,
\begin{equation}\label{eq:Stress tensor element}
P^{xy}=\sum_{i=1}^{N}\left[mv_{ix}v_{iy}-\frac{1}{2}\sum_{i\neq j}^{N}\frac{x_{ij}y_{ij}}{r_{ij}}\frac{\partial V(r_{ij})}{\partial r_{ij}}\right],
\end{equation}
with $N$ being the number of particles and $r_{ij} = |\mathbf{r_{i}-r_{j}}|$.

In Eq. \ref{eq:Green-Kubo}, $C(t)$ is the stress autocorrelation function (SACF) of particles. 
In practice, in Eq. \ref{eq:Green-Kubo}, the upper limit in the integral is limited by the cut-off time, which is defined approximately by the ratio of the simulation box length to the sound speed \cite{PhysRevE.79.026401}. 
Therefore, for a given number of particles, the accuracy of the EMD based calculations depends on the behavior of the SACF at long times.
Moreover, for 2D Yukawa systems, the SACF decays slower than $t^{-1}$ and faster than $t^{-1}$  with time at small and large values of the coupling parameter, respectively  \cite{PhysRevE.79.026401}.
The former means diverging integral in Eq. (\ref{eq:Green-Kubo}) at small values of the coupling parameter. Nevertheless, it turned out that Eq. (\ref{eq:Green-Kubo}) with a large enough  cut-off time gives  meaningful results for the  shear viscosity \cite{PhysRevE.84.046412, Liu}. The same is true for the diffusion coefficient computed from the Green-Kubo relation connecting the diffusion coefficient with the velocity autocorrelation function of particles \cite{PhysRevE.99.013203}. 

We use the EMD data to validated general features regarding dependence of the shear viscosity on the coupling and screening parameters.

\begin{figure}
\includegraphics[width=0.45\textwidth]{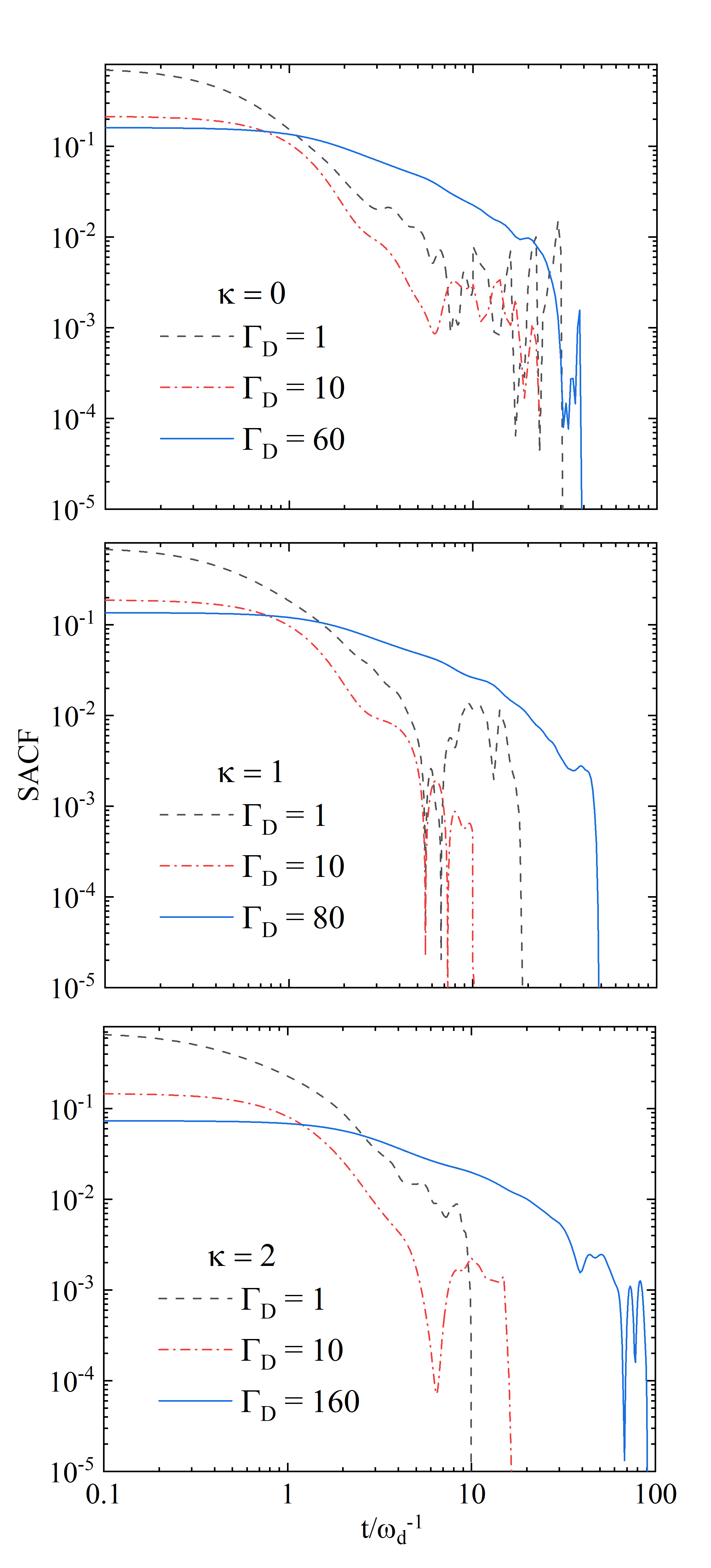}
\caption{ The SACF results at $\kappa=0$ (the top panel), $\kappa=1$ (the middle panel), and $\kappa=2$ (the bottom panel) for $\Gamma_D=1$, $\Gamma_D=10$, and $\Gamma_D\approx \Gamma_{m}$.
\label{fig:9}}
\end{figure}

\subsection{Simulation parameters}\label{s:sim_par}
We consider a system of particles enclosed in a square with periodic boundary conditions.
Particles with pair interaction potentials (1) and (2) are simulated using molecular dynamics, where streams of particles  are introduced as described in the previous subsection.   
The side length of the simulation box is defined by the number of particles as $L/a=\sqrt{\pi N}$,  with $a$ being the average distance between particles.
We consider N = 1024 particles. 
In our simulations, the length is considered in units of $a$ and the time in units of the inverse dipole frequency $1/\omega_d=({p_d^2}/(2\pi\epsilon_0ma^5))^{-1/2}$ \cite{Golden1, Golden2} (which is introduced in analogy with the plasma frequency, but does not describe a real collective oscillation mode); with $p_d$ being the electric dipole moment.
% , and $\eps=p^2/(4\pi\epsilon_0a^3)$ is used as the unit of energy. 
The system is characterized by 2 dimensionless parameters: first is coupling parameter $\Gamma_D$; second is the screening parameter $\kappa$.
The value of viscosity is given in units of $\eta_0=mn\omega_da^2$. The reduced shear rate is defined as $\gamma^*=(dv_x/dy)(1/\omega_d)$ and velocity values are presented in units of $v_0=a\omega_d$.

We study the dependence of shear viscosity on the coupling parameter $\Gamma_D$ considering $\kappa=0$, $\kappa=1 $, and $\kappa=2 $. The case $\kappa=0$ corresponds to a bare dipole-dipole interaction with the pair potential (1). For different $\kappa$ values, a system crystallizes at different $\Gamma_D$; therefore, for systems with different inverse screening lengths, the coupling parameter varied within different limits. The results presented in the  following sections are measured and averaged over 1000$\omega_d^{-1}$.
As a cross-validation of our NEMD code,  we have reproduced data for the shear viscosity of 2D Yukawa system reported by Donko et al \cite{Donko}. The corresponding comparison  of our results with that of Donko et al is shown in Appendix.

Using the EMD method, we calculated the shear viscosity for three values of the coupling parameter $\Gamma_D = 1, ~10$ and at $\Gamma_D\approx \Gamma_{m}$. The coupling parameter values, $\Gamma_{m}$, corresponding to the melting (crystallization) point for different screening parameters  were  reported by  Aldakul et al \cite{Aldakul}.
% The Fig. \ref{fig:9}. It is qualitatively clear that the integrand is convergent, so the integral itself must also converge (\ref{eq:Green-Kubo}). The SACFs cut-off time themselves are chosen based on the statistical noise that appears when calculating SACF, in view of the small number of particles and a small sample for averaging.
 The number of particles in our EMD simulations was set to $N=10^4$. The SACF data was averaged over 20 independent simulations. The SACF is presented in the units of $C_0=\eta_0\omega_d k_BTS$.
 %, and the stress tensor %components are calculated in the units of energy $\varepsilon = {p_d^2}/{(4\pi\epsilon_0 a^3)}$.

In the case of the bare dipole-dipole interaction,
a direct summation of the interaction force in MD is known to be highly time-consuming and inefficient due to scaling as $O(N^2)$ with respect to the number of particles.  To avoid 
this problem and reduce scaling to $O(N)$, the gradient-shifted force (GSF) electrostatics \cite{Lamichhane} based on the Wolf method   \cite{Wolf} is used.  Within the GSF 
electrostatics,  we set 
cutoff value and dumping coefficient to  $r_c/a=12a$ and  $\alpha=0.2a^{-1}$, respectively. These values allow to find converged data for structural and dynamical properties of 2D dipole systems \cite{Aldakul}.

\section{Results}\label{s:3}

\subsection{Shear thinning effect}
To begin with, in Fig. \ref{fig:2} we show the velocity distribution of particles along $y$ axis, i.e. perpendicular to the direction of flows, at different values of the period of momenta swap $\tau$ between slabs A and B. 
The results presented in Fig. \ref{fig:2} are for weakly correlated system with $\Gamma_D=1$ and with $\kappa=0$ (the left panel), $\kappa=1$ (the middle panel), and $\kappa=2$ (the right panel). The presented data is for the momenta exchange period $\omega_d\tau = 1$, 5, and 10.  From Fig. \ref{fig:2} we clearly observe that the more often the permutation of moments occurs, the greater the shear rate $\partial v_x/\partial y$. 
Further, as expected, the distribution of velocities between slabs A and B is a linear function of the distance between these slabs.

In Fig. \ref{fig:3}, we present  the velocity distribution of particles in the case of the strongly correlated system with $\Gamma_D=30$.
As in the weakly correlated case, we observe an increase in the shear rate with a decrease in the period of momentum exchange in the slabs.
Furthermore, the velocity distribution between streams remains to be a linear function of the distance. These behavior is general up to a crystallization point and allows to compute shear viscosity in a wide range of coupling parameter values using the NEMD method.  

% To show the velocity profiles $v_x(y)$, systems with the same $\Gamma_D$ and $\kappa$ parameters, but the various frequency of momentum exchange are counted. For systems with the same degree of screening $\kappa$, the following coupling parameters $\Gamma_D=1$, $\Gamma_D=30$ are selected. In all cases, the frequency of momentum exchange $\omega_d\tau = 1, 5 ,10$ are chosen. As can be seen in Fig.3 and Fig.4, the more often the momentum are changed, the higher the peaks, which means that the slope coefficient of the straight line will become greater.

The results presented in Figs. \ref{fig:2} and \ref{fig:3} have been measured  after the system is reached the stationary (equilibrium) regime.
The change in time of the shear rate, $\gamma$, at $\tau \omega_d=10$ is presented in Fig. \ref{fig:4} a)  and Fig. \ref{fig:4} b) for $\Gamma_D=1$ and    $\Gamma_D=30$, respectively.
From these figures we observe that to obtain a physical valid viscosity, one need to model for a long enough time so that the angle of inclination becomes approximately constant. For example, at momentum exchange period $10~\omega_d^{-1}$, it took about $60000~\omega_d^{-1}$ for the bare dipole system and  about $45000~\omega_d^{-1}$ for the screened dipole system to reach a stationary regime. 
Therefore, a system with a stronger inter-particle correlation takes longer to reach a steady state.
 In Fig. \ref{fig:4} c), the dependence of the the shear rate on the momenta exchange period $\tau$ is shown. As expected, one can observe an increase in the shear rate with a decrease in the period of momentum exchange in the slabs.

To obtain a physically correct shear viscosity, i.e. which is independent of shear rate,  it is necessary to permute the momentum as rarely as possible. 
At considered parameters, the optimal value of the frequency of momentum exchange is found to be once every 10 $\omega_d^{-1}$ time period.
If one increases the frequency of the momentum exchange, then shear viscosity can become a function of shear rate.
In Fig. \ref{fig:5}, the shear viscosity at different values of the momentum permutation period is presented for $\Gamma_D=1$ and $\Gamma_D=30$. 
From Fig. \ref{fig:5}, we clearly see that at $\Gamma_D=1$ (independent of screening parameter), the shear viscosity is approximately independent of the momentum exchange period at  $\tau \omega_d\gtrsim 10$. In contrast, at $\Gamma_D=1$ and $\tau \omega_d< 10$, we observe that the shear viscosity decreases as the momentum permutation period decreases.

Fig \ref{fig:6} shows the dependence of the viscosity on the shear rate at $\Gamma_D=1$ and $\Gamma_D=30$. From  Fig \ref{fig:6},  one can observe that the viscosity value approaches an equilibrium value  with the decrease in the shear rate. For $\Gamma_D = 30$, the change of the momentum permutation period in the range $10^{-1}\leq\tau \omega_d\leq 20$ does not lead to large enough change  in  the shear rate.
As the result,  we do not observe strong impact of the  shear rate variation on the viscosity value at $\Gamma_D=30$. 
% It can be assumed that as the coupling parameter increases, the dependence of the viscosity on the shear rate disappears. 

% To be sure, we calculated for all Gamma values with a momentum exchange period equal to $10 \omega_d \tau$, which corresponds to small shear rates. In fact, Fig. \ref{fig:5} and Fig. \ref{fig:6} show the same thing, the difference is only in the inverse dependence of the values along the horizontal axis between each other.

The decrease in the momentum permutation period is equivalent to the increase  in the shear rate as it was demonstrated in Figs. \ref{fig:2} and \ref{fig:3}.
The effect of the reduction of the shear viscosity with the increase in the shear rate is called shear thinning effect.
% Interestingly, we can see from Fig. \ref{fig:5} and Fig. \ref{fig:6}, that the shear viscosity remains nearly constant at all considered values of the momentum exchange period $0.1\leq \tau \omega_d \leq 20$ at $\Gamma_D=30$.
Thus, we are able to  observe from Fig. \ref{fig:5} and Fig. \ref{fig:6} the effect of shear thinning in a two-dimensional system of particles interacting through the repulsive dipole potential in the weakly correlated regime (e.g., at $\Gamma_D=1$).
One can expect to observe the shear thinning effect at $\Gamma_D=30$ in the case $\tau \omega_d\ll 10^{-1}$, but it is seems to be rather unrealistic limit in which the particles dynamics is strongly disturbed at the length scale of the mean inter-particle distance.  Therefore, next we focus on physically meaningful results on the shear viscosity values in the limit of low shear rates. 

% The dependence of the shear viscosity on the frequency of momentum exchange at coupling parameters $\Gamma_D$ = 1 and $\Gamma_D$ = 30 is clearly shown in Fig.7. It is interesting to note that with a small value of $\Gamma_D$ = 1, the more often the velocities are replaced to create flows, the lower the viscosity value, and with sufficiently rare substitutions, the viscosity reaches the true value. This is due to the fact that at low $\Gamma_D$ values, the thermal energy of particles and the potential energy of interaction are comparable, which contributes to a decrease in viscosity with frequent momentum exchange. But with sufficiently large values of coupling parameter ($\Gamma_D \geq$ 30), the thermal part makes a smaller contribution and the potential component prevails, which leads to a violation of the relationship between viscosity and the frequency of momentum exchange. As can be seen from results a shear- thinning effect is revealed for the dipole systems at low coupling parameters, in our case for  $\Gamma_D$ = 1, as for the Yukawa systems \cite{Hartmann}.

\subsection{Shear viscosity}
Let us consider shear viscosity in the limit of low shear rates in more detail.
For the screening parameters considered in this work, the phase transition point lie in the intervals $\Gamma_{m}$ = 67 $\pm$ 4 at $\kappa=0$, $\Gamma_m$ = 86 $\pm$ 6 at $\kappa=1$, and in the range $\Gamma_m$ = 163 $\pm$ 13 at $\kappa=2$. We varied the coupling parameter of the system from $\Gamma_D=1$ to approximately $\Gamma_D=\Gamma_m$. 

From Fig. \ref{fig:7} we see that the shear viscosity has a nonmonotonic dependence on the coupling parameter with a minimum at a moderate coupling value.
More specifically, the minimum of the shear viscosity occurs at $\Gamma_{min} \approx 10$ for the system with $\kappa=0$, at $10 \leq \Gamma_{min} < 20$ for the system with $\kappa=1$, and at $\Gamma_{min} \approx 30$ for the screened dipole system with $\kappa=2$. The change of the coupling parameter to lower or larger values results in the increase in the shear viscosity. 
This behavior is similar to that of observed for Yukawa systems and explained to be result of the competition between kinetic and correlation parts of the
pressure tensor \cite{PhysRevE.78.026408}. 

Additionally, from Fig. \ref{fig:7} we can observe that the minimum value of the shear viscosity decreases with the increase in the screening parameter.
Furthermore,  shear viscosity values at $\kappa=0$, $\kappa=1$, and $\kappa=2$ are approximately the same at $\Gamma_D=5$. 
At $\Gamma_D<5$, screening  leads to an increase in shear viscosity. In contrast, at $\Gamma_D>5$, screening  leads to a decrease in shear viscosity.

As sanity test of the NEMD results for 2D dipole systems, we performed calculations using  equilibrium MD data and  the Green-Kubo relation (\ref{eq:Green-Kubo}).
In Fig. \ref{fig:9}, we present results for the SACF at $\kappa=0$ (the top panel), $\kappa=1$ (the middle panel), and $\kappa=2$ (the bottom panel) for $\Gamma_D=1$, $\Gamma_D=10$, and $\Gamma_D\approx \Gamma_{m}$.
 $\Gamma_{ m}$ is the coupling parameter corresponding to the melting (crystallization) point from Ref. \cite{Aldakul}.
The general behavior of the SACF is decay with the increase in time. However, at $t>10 \omega_d^{-1}$, the SACF becomes strongly affected by noise due to finite number of particles in the main cell.  Here we used $N=10^4$ particles and averaged over independent $20$ simulations. For an accurate analysis of the behavior of the SACF at long times, one need much more particles in the main cell  \cite{PhysRevE.79.026401}. We have not explored this aspect of the SACF here, but we use the computed SACF results  to see if the NEMD data are adequate.

From Fig. \ref{fig:7}, we see that the EMD results computed using the SACF are in good agreement with the NEMD data with disagreement of about $4\%$ at $\Gamma_D=1$, $\kappa=0$ and $\kappa=1$. The largest disagreement of $8\%$ is observed for $\Gamma_D=10$ and $\kappa=1$.
In the other considered cases, the discrepancy between the NEMD and EMD data does not exceed a few percent.
These deviations are expected since the utility of the Green-Kubo relation for the viscosity calculations of 2D systems is problematic as it is discussed in Sec. \ref{s:2b}.
Nevertheless,  from Fig. \ref{fig:9}, we see that the EMD data has similar behavior as the NEMD data  with respect to the dependence on the coupling and screening parameters. 

The values of the shear viscosity computed using NEMD method  are given in Table \ref{tabl:tablichka}.
Additionally, for comparison, the shear viscosity results calculated from the EMD method are shown  in Table \ref{tabl:tablichka1}.

% Values of viscosity computed using EMD method using SACF are given in Table \ref{tabl:tablichka1}. SACF shown in Fig. \ref{fig:9}.
%[\textcolor{red}{Nasriddin: Show the shear viscosity values for $\kappa=0$, $\kappa=1$, and $\kappa=2$ in a Table !}]
\begin{table}[h]
\caption{The shear viscosity values in the limit of low shear rates for dipole systems with $\kappa=0$, $\kappa=1$ and $\kappa=2$,obtained by the NEMD method as described in Sec. \ref{s:2a}.}
\label{tabl:tablichka}

\begin{tabularx}{0.4\textwidth}{ | >{\centering\arraybackslash}X | >{\centering\arraybackslash}X | >{\centering\arraybackslash}X | >{\centering\arraybackslash}X |}
\hline
$\Gamma_D$&$\kappa=0$&$\kappa=1$&$\kappa=2$\\ \hline
1 & 0.61 & 0.64 & 0.71 \\
2 & 0.39 & 0.41 & 0.44 \\
5 & 0.27 & 0.27 & 0.28 \\
10 & 0.26 & 0.24 & 0.22 \\
20 & 0.31 & 0.26 & 0.20 \\
30 & 0.39 & 0.31 & 0.20 \\
40 & 0.50 & 0.36 & 0.21 \\
50 & 0.62 & 0.45 & 0.23 \\
60 & 0.85 & 0.54 & 0.24 \\
70 &  & 0.67 & 0.26 \\
80 &  & 0.86 & 0.28 \\
90 &  &  & 0.30 \\
100 &  &  & 0.33 \\
120 &  &  & 0.41 \\
140 &  &  & 0.51 \\
160 &  &  & 0.71 \\
\hline
\end{tabularx}
\end{table}

\begin{table}[h]
\caption{The shear viscosity values for dipole systems with $\kappa=0$, $\kappa=1$ and $\kappa=2$, obtained by the EMD method as described in Sec. \ref{s:2b}.}
\label{tabl:tablichka1}

\begin{tabularx}{0.4\textwidth}{ | >{\centering\arraybackslash}X | >{\centering\arraybackslash}X | >{\centering\arraybackslash}X | >{\centering\arraybackslash}X |}
\hline
$\Gamma_D$&$\kappa=0$&$\kappa=1$&$\kappa=2$\\ \hline
1 & 0.58 & 0.61 & 0.70 \\	
10 & 0.26 & 0.22 & 0.21 \\	
60 & 0.82 & - & - \\
80 & - & 0.87 & - \\
160 & - & - & 0.70 \\
\hline
\end{tabularx}
\end{table}
% Indeed, as can be seen on the Fig. \ref{fig:8}, the shear viscosity value increases. 
% The greater the degree of screening of the system, the further along the $\Gamma_D$ axis and lower along the $\eta/\eta_0$ axis the minimum of the function shifts.  This confirms the fact that for systems with a lower $\kappa$, the crystallization moment shifts to the range of lower $\Gamma_D$, and vice versa for large $\kappa$ values, the crystallization moment shifts to the range of higher $\Gamma_D$. 

\begin{figure} 
\includegraphics[width=0.5\textwidth]{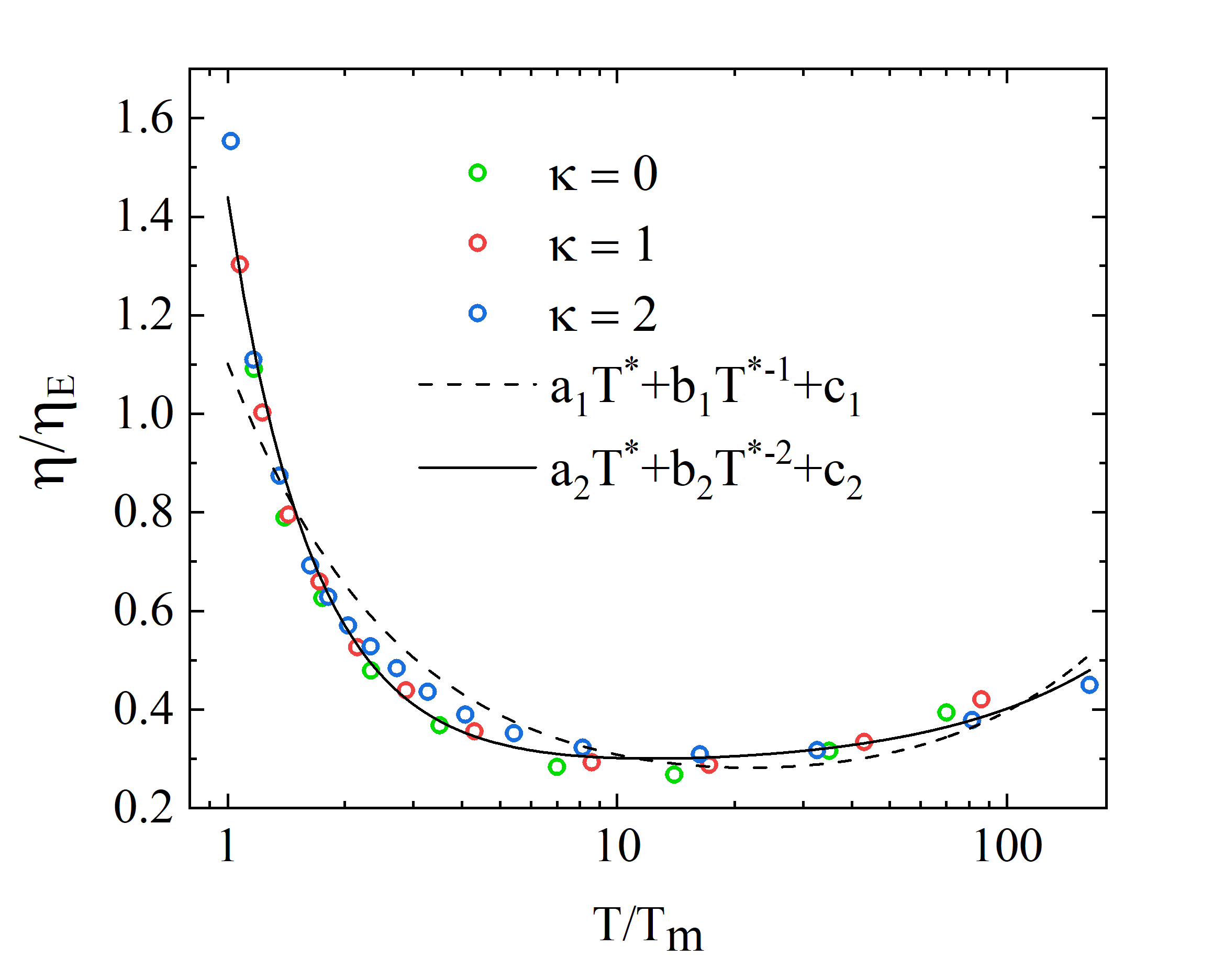}
\caption{ The dependence of the shear viscosity on the reduced temperature $T/T_m$, where the shear viscosity is presented in the units of $\eta_E$. The dash line is the best fit obtained using Eq. (\ref{eq:eta1}); the solid line is the best fit based on Eq. (\ref{eq:eta2}). The data was fitted using the method of least squares.
\label{fig:8}}
\end{figure}

\subsection{Universal scaling law}

It is known that one can express the viscosity dependence on temperature via some universal scaling low \cite{PhysRevResearch.4.033064}. 
For example, in the case of 2D Yukawa systems, one can find such expression by expressing the viscosity value in the units of $\eta_E = mn\omega_E a^2$ and as the function of the reduced temperature $T^*=T/T_m$, where $\eta_E$ is the Einstein frequency and $T_m$ is a melting temperature. 

We have performed analysis of the computed viscosity data using $\eta_E$.
For that, the Einstein frequency values at different parameters were calculated as  \cite{Saigo}:
\begin{equation}\label{eq:Einstein frequency}
\omega_E^2 = \frac{1}{3m}\sum_{i\neq j}\Delta V(\vec{r_i}-\vec{r_j}).
\end{equation}

The $\eta/\eta_E$ dependence on $T/T_m$ is presented in Fig. \ref{fig:8} for $\kappa=0$, $\kappa=1$, and $\kappa=2$.
From Fig. \ref{fig:8}, one can clearly see that that there is some universality in thevdependence of $\eta/\eta_E$  on $T/T_m$, which is not sensitive  to the screening parameter. 

In the case of  2D Yukawa systems, the  universal scaling law for $\eta/\eta_E(T/T_m)$ reads \cite{Hartmann}:
\begin{equation}\label{eq:eta1}
    \eta/\eta_E = a_1T^*+ b_1T^{*-1}+c_1,
\end{equation}
where $a_1$, $b_1$, and $c_1$ are fitting parameters and $T^*=T/T_m$ is reduced temperature.

First, we checked if Eq. (\ref{eq:eta1}) can describe  2D dipole systems. 
The best fit obtained using Eq. (\ref{eq:eta1}) and the method of least squares is shown in Fig. \ref{fig:8} by dashed line, where $a_1 = 0.00187897$, $b_1 = 0.9$, and $c_1 = 0.19964258$.
Clearly, Eq. (\ref{eq:eta1}) is not able to provide an adequate universal scaling low.
Instead, we found that much better description is provided by replacing $T^{*-1}$ term with $T^{*-2}$, i.e., by using
\begin{equation}\label{eq:eta2}
    \eta/\eta_E = a_2T^*+ b_2T^{*-2}+c_2,
\end{equation}
where $a_2$, $b_2$, and $c_2$ are fitting parameters.

The best fit based on Eq. (\ref{eq:eta2}) is shown in Fig. \ref{fig:8} using solid line, where $a_2 = 0.00124$, $b_2 = 1.16076$, and $c_2 = 0.27754$. From Fig. \ref{fig:8}, we observe that Eq. (\ref{eq:eta2}) provides an adequate universal scaling low for 2D repulsive dipole systems.

% \textcolor{red}{
% It turned out that the viscosity of two-dimensional dipole systems, as well as the viscosity of two-dimensional systems \cite{Hartmann}, obeys a universal law, as can be seen in Fig. \ref{fig:8}. Viscosity is dimensionless by $\eta_E = mn\omega_E a^2$, where $\eta_E$ is Einstein frequency. The physical meaning of this characteristic of the system is as follows: it is the value of the oscillation frequency of the structural units of the substance, averaged over the number of particles and time. The Einstein frequency depends on the $\kappa$ and was calculated using the following formula \cite{Saigo}, which used for fcc lattice:
% \begin{equation}\label{eq:Einstein frequency}
% \omega_E^2 = \frac{1}{3m}\sum_{i\neq j}\Delta V(\vec{r_i}-\vec{r_j})
% \end{equation}
% On the abscissa axis, the dimensionless temperature $T/T_m = \Gamma_{D,m}/\Gamma_D$, where $T_m$ is a melting temperature. The data obtained with the help of these normalizations obey the universal law. It can be argued that the universal law is satisfied for $\kappa$ values in the interval from 0 to 2. Curve that fits this data $\eta^* = a_2T^*+ b_2T^{*-2}+c_2$, where $a_2 = 0.00124, b_2 = 1.16076, c_2 = 0.27754$. Also in Fig. \ref{fig:8} shows how these data correspond to the fitting curve for Yukawa systems. It can be seen that the corresponding universal law for Yukawa systems does not fit well for the data for dipole systems.}

\section{Conclusion}
The  shear viscosity  of two-dimensional  dipole and screened dipole systems is investigated using the NEMD. 
The optimal values of the momentum exchange frequency and equilibration time are analyzed for computing the shear viscosity values at different coupling and screening parameters.
The dependence of the shear viscosity  on the coupling  parameter $\Gamma$ in the limit of low shear rates for different screening parameters is presented. 
It is found that screening leads to an increase in the shear viscosity at $\Gamma_D<5$. In contrast, at $\Gamma_D>5$ the shear viscosity of 2D dipole systems decreases with increasing screening parameter. 
As expected, the shear viscosity of 2D dipole systems has a minimum at intermediate coupling parameters. The value of the coupling parameter corresponding  to the minimum of the shear viscosity shifts to larger values with an increase in the screening degree. Furthermore,  a shear thinning effect is revealed for 2D dipole systems at low values of the coupling parameter.

Our extensive NEMD simulations have allowed us, to the best of our knowledge, to calculate the shear viscosity of classical 2D repulsive dipole systems for the first time. Furthermore, we found a simple  fitting curve which provides single universal scaling law valid for both the bare dipole - dipole pair interaction potential and  the screened dipole-dipole pair interaction potential.
Taking into account the relevance of dipole systems for various fields of physics and the general interest from the point of view of statistical physics, we believe that the presented study is a valuable addition to the physics of strongly correlated 2D systems.

% These studies, in fact, are a logical continuation of the study of the shear viscosity of two-dimensional Yukawa systems, since even the behavior of the viscosity function itself is similar in structure to Yukawa systems. Having an understanding of what value the viscosity takes for a given value of the coupling parameter and the degree of screening, one can objectively assess how "coupled" the system is compared to the system with other parameters $\Gamma$ and $\kappa$.

\appendix

\section*{Appendix: The shear viscosity of classical 2D Yukawa systems}

% In this section, the correctness of the code and the implementation of the non equilibrium reverse molecular dynamics method are discussed.
In order to verify the correctness of our  implementation of the NEMD method, we computed the shear viscosity of 2D Yukawa systems  in the limit of low  shear rates and validated our results by comparison with previously published data by Donko et. al.  \cite{Hartmann}. 

In the case of the Yukawa system, the pair interaction potential is defined as $U=Q^2\exp(-r/\lambda_D)/4\pi\eps_0r$. 
The Yukawa system is characterized by the following dimensionless parameters: $\Gamma$ = $Q^2/2\pi\eps_0ak_BT$ and $\kappa$ = $a/\lambda_D$, $a$ = $(1/\pi n)^{1/2}$, $a$ - Wigner-Seitz radius, $\lambda_D$  - Debye screening length, $n$ is the number  density of particles, $\omega_p$ = $(Q^2/2\pi\eps_0ma^3)^{1/2}$,  $\omega_p$  is the 2D analog of the plasma frequency, and $\eta^*$ = $\eta/mn\omega_pa^2$. 

\begin{figure} 
\includegraphics[width=0.45\textwidth]{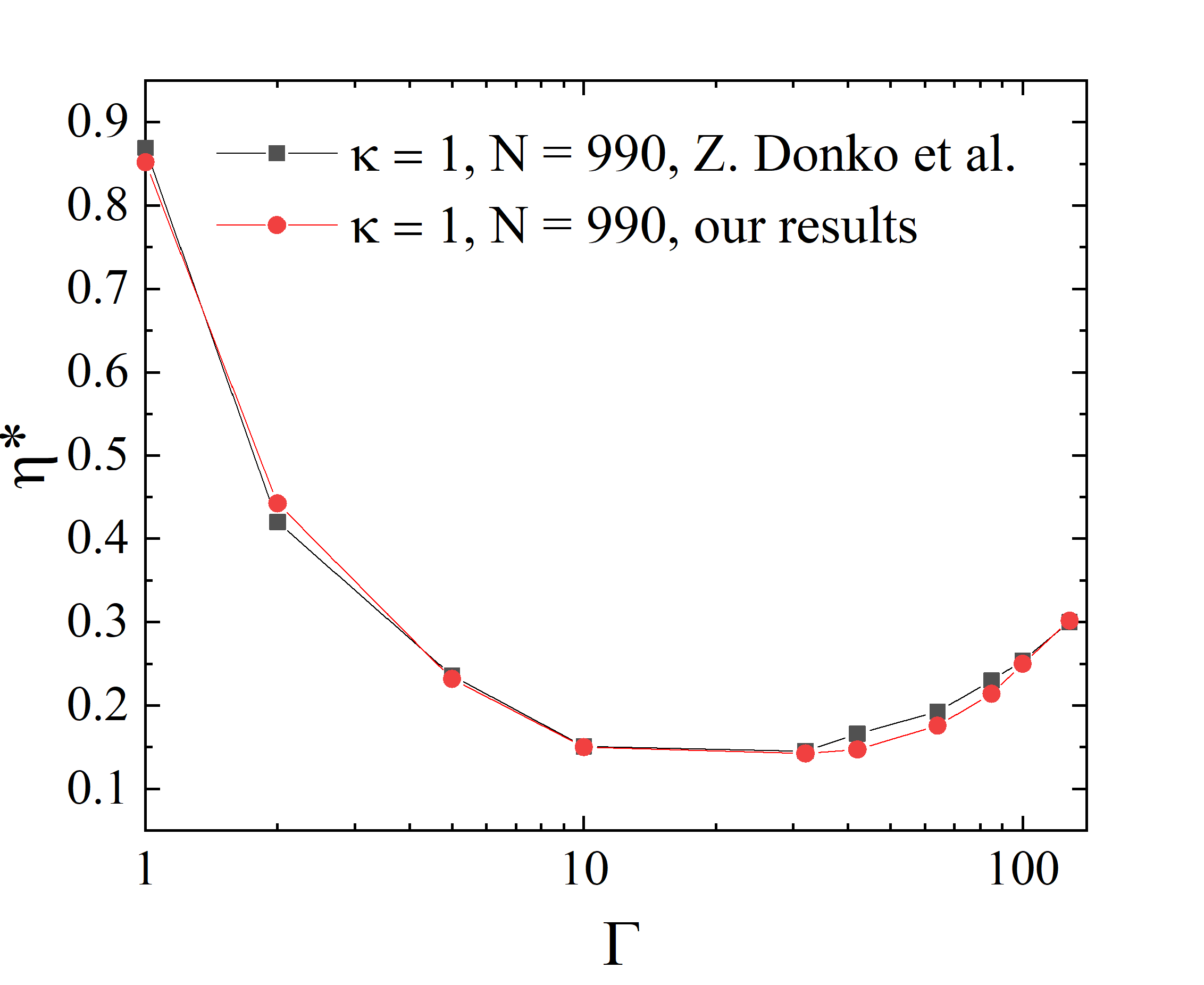}
\caption{  Shear viscosity in the limit of low shear rates for the 2D Yukawa system. The comparison with the data from Ref. \cite{Hartmann} is presented.
\label{fig:10}}
\end{figure}

The comparison of our results with the data from Ref. \cite{Hartmann} is presented in Fig. \ref{fig:10} for $\kappa=1$. 
As can be seen from  Fig. \ref{fig:10}, our results  have a fairly accurate agreement with the data from Ref. \cite{Hartmann}. 
% Therefore, all of the above can be considered reliable data.

\section*{Acknowledgments}
This research is funded by the Science Committee of the Ministry of Education and Science of the Republic of Kazakhstan (Grant AP08855651).
% %%%%%%%%%%%%%%%%%%%%%%%%%%%%%%%%%%%%%%%%%%%%%%%%%%%%%%%%%%%%%%%%%%%%%%%%%%%%%%% 
% Bibliography
% %%%%%%%%%%%%%%%%%%%%%%%%%%%%%%%%%%%%%%%%%%%%%%%%%%%%%%%%%%%%%%%%%%%%%%%%%%%%%%%%

%  \bibliographystyle{apsrev4-1}
 \bibliography{ref}

%\section*{Acknowledgments}
 %  This research has been funded by the Science Committee of the Ministry of Education and Science of the Republic of Kazakhstan (Grant No AP08855651). 
 %\section*{Data Availability}  The data that support the findings of this study are available from the corresponding author upon reasonable request.

\end{document}